\documentclass[11pt]{article}
\usepackage[dvips]{epsfig}

 \advance \textwidth by -2in
 \advance \textheight by -3in
\def\##1{\underline{#1}}
\def\=#1{\underline{\underline{#1}}}

\def\+#1{\underline{\bf #1}}
\def\*#1{\underline{\underline{\bf #1}}}

\def\eps{\epsilon}

\def\.{\mbox{ \tiny{$^\bullet$} }}

\def\le{\left(}
\def\ri{\right)}

\def\lec{\left\{}
\def\ric{\right\}}

\def\c#1{\cite{#1}}
\def\l#1{\label{#1}}
\def\r#1{(\ref{#1})}

\begin{document}
\noindent Submitted for publication in {\it Microwave \& Optical Technology Letters}
\vskip 0.4cm

\noindent {\bf NEGATIVE PHASE VELOCITY IN ISOTROPIC
DIELECTRIC--MAGNETIC MEDIUMS VIA HOMOGENIZATION} \vskip 0.2cm

\noindent  {\bf Tom G. Mackay$^1$ and Akhlesh Lakhtakia$^2$}
\vskip 0.2cm

\noindent {\sf $^1$ School of Mathematics\\
\noindent University of Edinburgh\\
\noindent Edinburgh EH9 3JZ, United Kingdom}
\vskip 0.4cm

\noindent {\sf $^2$ CATMAS~---~Computational \& Theoretical Materials Sciences Group \\
\noindent Department of Engineering Science \& Mechanics\\
\noindent 212 Earth \& Engineering Sciences Building\\
\noindent Pennsylvania State University, University Park, PA
16802--6812} \vskip 0.4cm

\noindent {\bf ABSTRACT:} We report on a strategy for achieving
negative phase velocity (NPV) in a homogenized composite medium
(HCM) conceptualized using the Bruggeman formalism. The constituent material
phases of the HCM do not support NPV propagation. The HCM and its constituent
phases are isotropic dielectric--magnetic mediums; the real parts
of their permittivities/permeabilities are negative--valued whereas the real
parts of their permeabilities/permittivities are positive--valued.


\vskip 0.4cm

\noindent{\bf 1. INTRODUCTION}

A plane wave is said to propagate with negative phase velocity
(NPV) if its phase velocity is projected opposite to the
time--averaged Poynting vector. A host of exotic electromagnetic
phenomenons follow as consequence of NPV, most notably negative
refraction \c{Pendry04}, as is well--documented elsewhere
\c{LMW_CM,Rama}.

In the absence of  readily available,  naturally occurring
materials which support NPV propagation\footnote{We note the
possibility of NPV induced by the effects of (a) special
relativity in simple dielectric--magnetic mediums \c{ML_STR} and
(b) general relativity in vacuum \c{MLS_GTR}.},
 the realization of artificial \emph{metamaterials} which are effectively homogeneous
 and which
support NPV propagation has been the focus of considerable
attention \c{Rama}. NPV metamaterials for performance in the microwave regime have
been realized \c{SSS}--\c{NPV_expt3}, and progress towards the same
goal in the optical regime continues to be made.

The simplest  medium which supports NPV propagation is the
idealization  represented by the nondissipative isotropic
dielectric--magnetic medium with relative permittivity $\eps < 0$
and relative permeability $\mu < 0$. In reality, the effects of
dissipation necessitate that $\eps$ and $\mu$ are complex--valued.
In a dissipative isotropic dielectric--magnetic medium, NPV is
indicated by the satisfaction of the inequality \c{LMW_CM,DL04}
\begin{equation}
 \frac{\mbox{Re} \, \lec \eps \ric}{\mbox{Im} \, \lec \eps \ric}
+ \frac{\mbox{Re} \, \lec \mu \ric}{\mbox{Im} \, \lec \mu \ric}  <
0\,. \l{NPV_em}
\end{equation}
In this communication we address the question:  can the NPV
condition \r{NPV_em} be satisfied by a homogenized composite
medium (HCM) which arises from constituent material phases which do not
themselves support NPV propagation?

\noindent{\bf 2. HOMOGENIZATION}

Let us consider the homogenization of two constituent material phases: phase
$a$ and phase $b$. Both material phases are taken to be isotropic
dielectric--magnetic mediums with relative permittivities
$\eps_{a,b}$ and relative permeabilities $\mu_{a,b}$. The relative
permittivity and relative permeability of the  HCM are written as $\eps_{HCM}$ and
$\mu_{HCM}$, respectively. In accordance with the principle of
causality and because of the implicit time--dependence $\exp(-i\omega t)$, we have
\begin{equation}
\left.
\begin{array}{l}
 \mbox{Im} \, \lec \eps_{\ell} \ric > 0
\\
 \mbox{Im} \,
\lec \mu_{\ell} \ric > 0
\end{array}
\right\}, \qquad (\ell = a, b, HCM)\,\l{causality}
\end{equation}
 as the mediums under consideration are
assumed to be passive.
 The
volume fraction of phase $\ell$ is denoted by $f_\ell \in (0,1)$
($\ell = a,b$) with $f_a + f_b = 1$.

Conventional approaches to homogenization of particulate materials, such as provided by the
Bruggeman and the Maxwell Garnett formalisms \cite{LOCM}, run into
difficulties within the context of NPV--supporting HCMs. These
formalisms have been shown to be inappropriate if $\mbox{Re} \, \lec \eps_a \ric
\, \mbox{Re} \, \lec \eps_b \ric < 0$ or $\mbox{Re} \, \lec \mu_a
\ric \, \mbox{Re} \, \lec \mu_b \ric < 0$,
 at least in the weakly dissipative regime \c{ML_Limitations}.
  There are no such
 difficulties provided that
 $\mbox{Re} \, \lec \eps_a \ric
\, \mbox{Re} \, \lec \eps_b \ric > 0$ and $\mbox{Re} \, \lec \mu_a
\ric \, \mbox{Re} \, \lec \mu_b \ric > 0$. However, in view of
\r{causality}, it is clear that the NPV condition \r{NPV_em}
cannot be satisfied if $\mbox{Re} \, \lec \eps \ric > 0$ and
$\mbox{Re} \, \lec \mu \ric > 0$. We therefore explored  the
prospects of achieving NPV in a HCM arising from components with
either
\begin{itemize}
\item[ ] {\em Case I:\/} $\mbox{Re} \, \lec \eps_{a,b} \ric < 0$ and
$\mbox{Re} \, \lec \mu_{a,b} \ric > 0$,
or
\item[ ] {\em Case II:\/} $\mbox{Re} \, \lec \eps_{a,b} \ric > 0$ and
$\mbox{Re} \, \lec \mu_{a,b} \ric < 0$.
\end{itemize}
Cases I and II are duals of each other, and only one of the two needs
further investigation.

The relative permittivity and relative permeability of the
corresponding HCM are estimated by the Bruggeman formalism as
\c{M_CM}
\begin{equation}
\left.
\begin{array}{l}
 \eps_{HCM} = \displaystyle{\frac{f_a \eps_a \le \eps_b + 2 \eps_{HCM} \ri +
f_b \eps_b \le \eps_a + 2 \eps_{HCM} \ri }{f_a  \le \eps_b + 2
\eps_{HCM} \ri + f_b \le \eps_a + 2 \eps_{HCM} \ri}}\,\\ \vspace{-12pt} \\
 \mu_{HCM} = \displaystyle{\frac{f_a \mu_a \le \mu_b + 2 \mu_{HCM} \ri +
f_b \mu_b \le \mu_a + 2 \mu_{HCM} \ri }{f_a  \le \mu_b + 2
\mu_{HCM} \ri + f_b \le \mu_a + 2 \mu_{HCM} \ri}}\,
\end{array}
\right\} . \l{Br}
\end{equation}
As a representative example, let
\begin{equation}
\left.
\begin{array}{ll}
\eps_a = -6 + 0.9i, & \quad \mu_a = 1.5 + 0.2 i \\
\eps_b = -1.5 + i , & \quad \mu_b = 2 + 1.2 i
\end{array}
\right\}\,,
\end{equation}
in accordance with Case I.
The Bruggeman estimates   $ \eps_{HCM}$ and  $\mu_{HCM}$ are
plotted as functions of the volume fraction $f_a$ in Figures~1 and 2,
respectively. Whereas  $\mbox{Re} \, \lec \eps_{HCM}
\ric$ follows  an almost linear progression between its
constraining values of $\mbox{Re} \, \lec \eps_{b} \ric$ at $f_a =
0$ and $\mbox{Re} \, \lec \eps_{a} \ric$ at $f_a = 1$, and   similar
dependences are evinced by both $\mbox{Re} \, \lec \mu_{HCM} \ric$ and
$\mbox{Im} \, \lec \mu_{HCM} \ric$,
$\mbox{Im} \, \lec \eps_{HCM} \ric$ displays a markedly nonlinear
relationship with respect to $f_a$.

In Figure~3, the NPV parameter
\begin{equation}
\rho_\ell = \frac{\mbox{Re} \, \lec \eps_\ell \ric}{\mbox{Im} \,
\lec \eps_\ell \ric} + \frac{\mbox{Re} \, \lec \mu_\ell
\ric}{\mbox{Im} \, \lec \mu_\ell \ric}, \qquad (\ell = a, b, HCM)
\end{equation}
is graphed against the volume fraction $f_a$. For the constituent material phases
we have the constant values $\rho_a = 0.83$ and $\rho_b = 0.17$;
i.e., neither constituent phase supports NPV propagation. In
contrast, $\rho_{HCM}$ is negative--valued for $0.28 < f_a <
0.92$. Thus, we see that the HCM supports NPV propagation across a
wide range of volume fractions.

\noindent{\bf 3. CONCLUDING REMARKS}

In answer to the question posed in Section 1: an HCM which
supports NPV propagation, arising from constituent material phases
which do not support NPV propagation, may be conceptualized
through homogenizing components with $\mbox{Re} \, \lec \eps_{a,b}
\ric < 0$. In view of \r{NPV_em}, it may be inferred  that the
prospects for NPV propagation are increased through considering
constituent phases with relatively small $\mbox{Im} \, \lec
\eps_{a,b} \ric$ and relatively large $\mbox{Im} \, \lec \mu_{a,b}
\ric$.

Since the relative permittivities and relative permeabilities are
decoupled within the Bruggeman formalism, our demonstration with Case I
also holds for Case II.  However, we note that
in practice suitable materials with $\mbox{Re} \, \lec \eps_{a,b}
\ric < 0$ may be more readily available than those with $\mbox{Re}
\, \lec \mu_{a,b} \ric < 0$.

\newpage

\begin{figure}[!ht]
\centering \psfull \epsfig{file=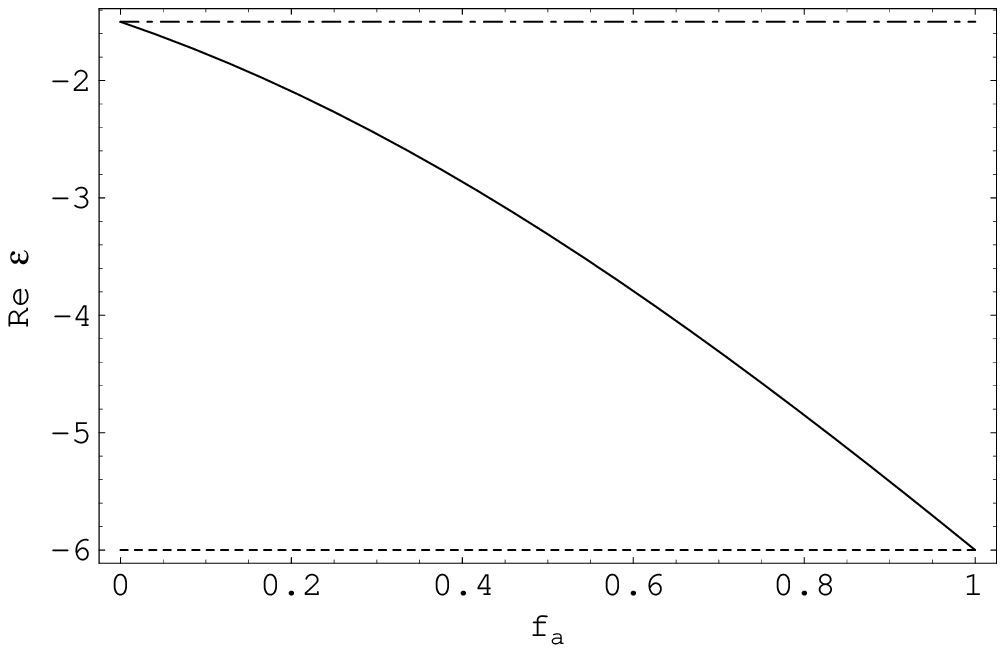,width=5.0in}
\epsfig{file=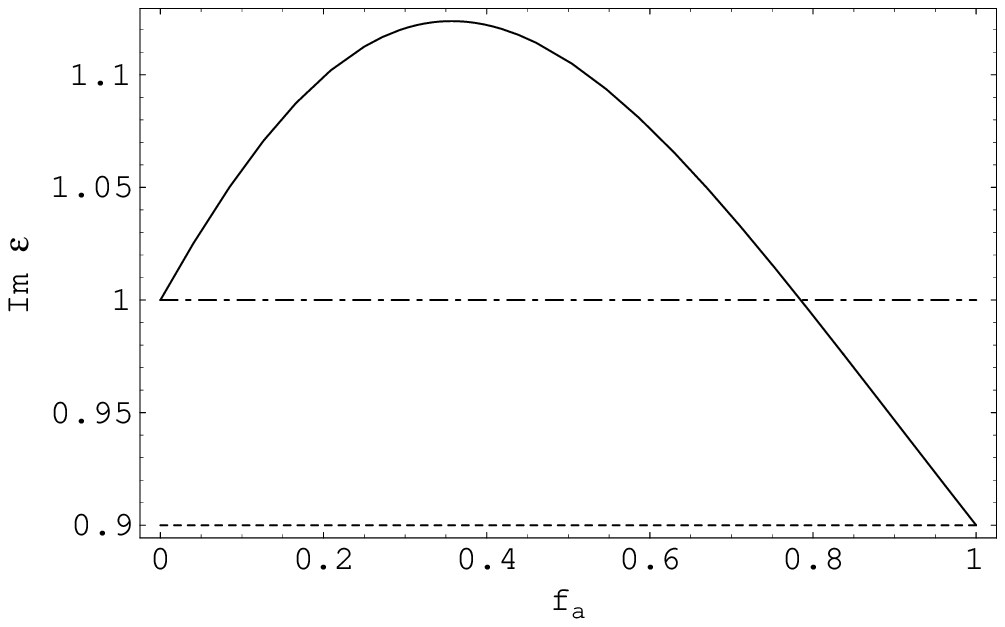,width=5.0in}
 \caption{\label{fig1} The real (top) and imaginary (bottom) parts
 of the relative permittivity $\eps_{HCM}$ of the HCM, as
 estimated using the Bruggeman formalism, plotted (solid curve) against volume
 fraction $f_a$. The dashed horizontal line represents $\eps_a$
 and
  the broken dashed horizontal line represents $\eps_b$.
  }
\end{figure}

\newpage

\begin{figure}[!ht]
\centering \psfull \epsfig{file=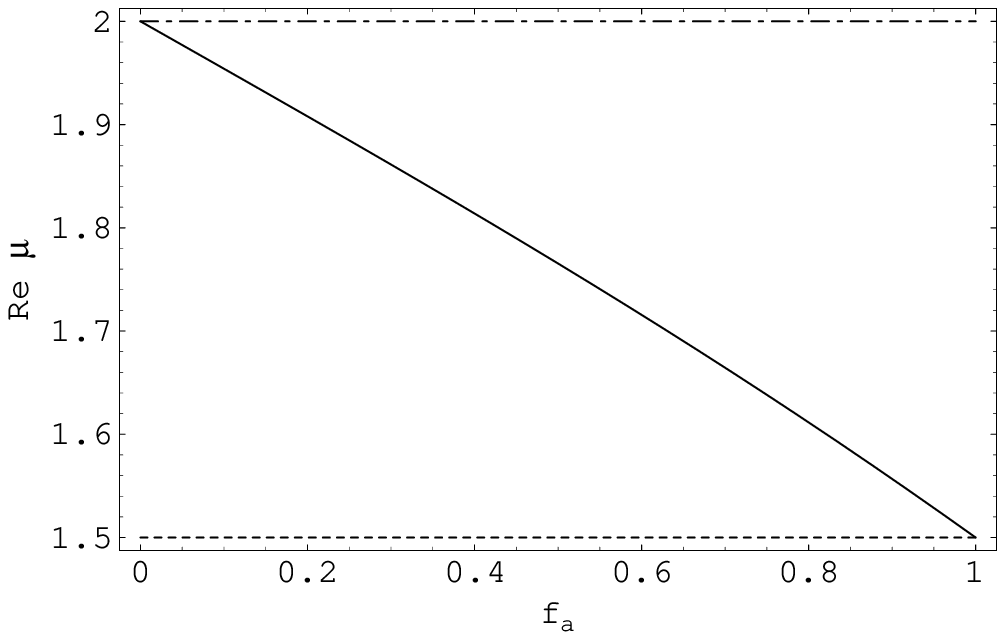,width=5.0in}
\epsfig{file=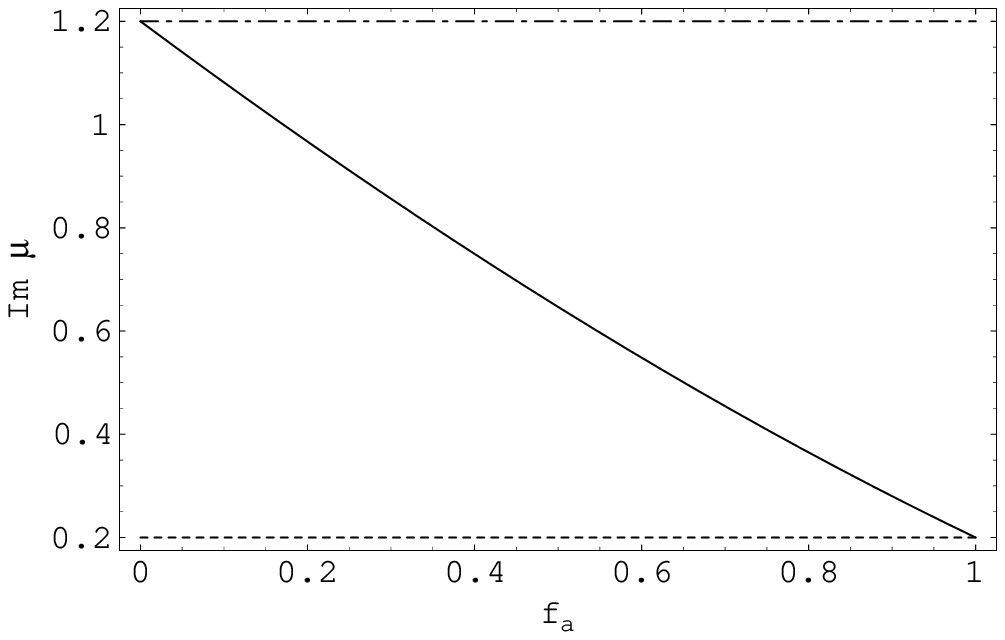,width=5.0in}
 \caption{\label{fig2} The
real (top) and imaginary (bottom) parts of the
  relative permeability $\mu_{HCM}$ of the HCM, as
 estimated using the Bruggeman formalism, plotted (solid curve) against volume
 fraction $f_a$. The dashed horizontal line represents $\mu_a$
 and
  the broken dashed horizontal line represents $\mu_b$.
  }
\end{figure}

\newpage

\begin{figure}[!ht]
\centering \psfull \epsfig{file=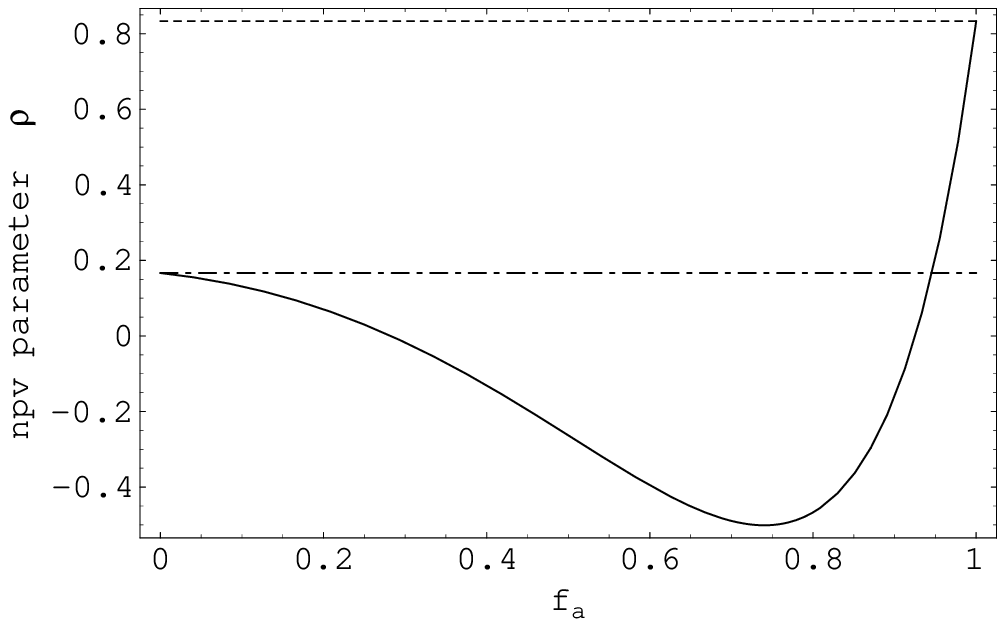,width=5.0in}
 \caption{\label{fig3} The NPV parameter $\rho_{HCM}$ of the HCM, as
 estimated using the Bruggeman formalism, plotted (solid curve)  against volume
 fraction $f_a$. The dashed horizontal line represents $\rho_a$ for phase $a$
 and
  the broken dashed horizontal line represents $\rho_b$ for  phase $b$.
  }
\end{figure}


\begin{thebibliography}{99}

\bibitem{Pendry04}
J.B. Pendry,
 Negative refraction,
Contemp Phys  45 (2004), 191--202.

\bibitem{LMW_CM}
A. Lakhtakia, M.W. McCall and W.S. Weiglhofer, Negative
phase--velocity mediums, In: W.S. Weiglhofer and A. Lakhtakia
(eds), Introduction to complex mediums for electromagnetics and
optics, SPIE Press, Bellingham, WA, USA, 2003.


\bibitem{Rama}
S.A. Ramakrishna, Physics of negative refractive index materials,
Rep Prog Phys 68 (2005), 449--521.

\bibitem{ML_STR}
T.G. Mackay  and A. Lakhtakia, Negative phase velocity in a
uniformly moving, homogeneous, isotropic, dielectric--magnetic
medium, J Phys A: Math Gen  37 (2004),  5697--5711.

\bibitem{MLS_GTR}
T.G. Mackay, A. Lakhtakia and S.  Setiawan, Gravitation and
electromagnetic waves with negative phase velocity, New J Phys 7
(2005), 75.

\bibitem{SSS}
R.A. Shelby, D.R. Smith and S. Schultz, Experimental verification
of a negative index of refraction, Science {292} (2001)  77--79.

\bibitem{NPV_expt1}
A. Grbic  and G.V. Eleftheriades,  Experimental verification of
backward--wave radiation from a negative index metamaterial, J
Appl Phys 92 (2002), 5930--5935.

\bibitem{NPV_expt2}
C.G. Parazzoli, R.B. Greegor, K. Li, B.E.C. Koltenbah and M.
Tanielian,
 Experimental verification and simulation of negative index of refraction
 using Snell's law,
Phys Rev Lett  90 (2003)  107401.

\bibitem{NPV_expt3}
A.A. Houck, J.B. Brock  and I.L. Chuang, Experimental observations
of a left--handed material that obeys Snell's law, Phys Rev Lett
90 (2003), 137401.

\bibitem{DL04}
R.A. Depine and A. Lakhtakia,
A new condition to identify isotropic dielectric--magnetic materials displaying negative phase velocity, Microwave   Opt  Technol  Lett  41 (2004) 315--316.

\bibitem{LOCM}
A. Lakhtakia
(ed), Selected papers on linear optical composite materials, SPIE
Optical Engineering Press, Bellingham, WA, USA, 1996.


\bibitem{ML_Limitations}
T.G.  Mackay and A. Lakhtakia, A limitation of the Bruggeman
formalism for homogenization, Opt Commun  234 (2004), 35--42.


\bibitem{M_CM}
T.G. Mackay, Homogenization of linear and nonlinear complex
composite materials, In: W.S. Weiglhofer and A. Lakhtakia (eds),
Introduction to complex mediums for electromagnetics and optics,
SPIE Press, Bellingham, WA, USA, 2003.

\end{thebibliography}
\end{document}